\documentclass[11pt,dvips]{article}

\begin{document}
\title{{\bf {\LARGE Neutron Scattering by Superfluid He II  
\\[-0.15cm] about Dispersion Minimum}}}
\author{  J. X. Zheng-Johansson
\\[0.1cm]
{\small  {\it Institute of Fundamental Physics Research,  611 93 Nyk\"oping, Sweden }}
\\[0.1cm]
{\small   October, 2004; updated August, 2006}
} 
\date{}

\maketitle
\def\w{\omega}
\def\ph{{\rm ph}}
\def\cph{{\rm cph}}
\def\rb{{\bf  r }}
\def\Rb{{\bf  R }}
\def\ub{{\bf  u }}
\def\l{ l}
\def\el{{\rm el}}
\def\inel{{\rm inel}}
\def\lf{\left}
\def\rt{\right}
\def\qb{{\bf  q}}
\def\eb{{\bf  e}}
\def\Kb{{\bf  K}}
\def\Ds{\Delta_{\rm s}}
\def\neu{{\rm ne}}
\def\Db{{\Delta_b}}
\def\Dbn{{\Delta_{b \nf}}}
\def\Dh{\Delta h}
\def\dn{\delta_{{\rm \nf}}}
\def\dv{\delta_{{\rm v}}}
\def\ds{\delta_{s}}
\def\da{\delta_{\alpha}}

\begin{abstract}
We derive the structure factor for superfluid He II about the energy dispersion minimum 1.93 1/\AA.
\\
\\
{\scriptsize  $^*$This e-print version (2) contains fuller list of the earlier reports on the new superfluid theory.} 
\end{abstract}

\section{Introduction}

For the superfluid described as consisting of a disordered collection of localized atoms\cite{Sup2,Sup1}, the total scattering function may be written in anology to that of a harmonic solid: 
$$ \displaylines{\refstepcounter{equation} \label{eq-Sqw1}
\hfill S(q,\w) 
= S(q)  + S_{\ph} +  S_{2 \ph} +\ldots   
 \hfill (\ref{eq-Sqw1})
}$$
There are of course basic distinctions between $S(q,\w) $ for the superfluid and for solids, as will  be commented where relevant. 
In the present article, within the framework of the microscopic theory of superfluid He II\cite{Sup2,Sup1}, we derive  an expression for the (static) structure factor, the first term $S(q)$ of (\ref{eq-Sqw1}), for the superfluid in the vicinity of $q_b=1.93$ 1/AA at which the measured phonon excitation spectrum $\w(q)$ presents a minimum.

$S(q)$ generally describes the zero-phonon scattering of neutrons, i.e.   the scattering processes which do not produce {\it harmonic} displacement of the superfluid atoms. Denoting the associated  atomic displacement in the absolute coordinate system by $u(q)$, then $ u(q_b)=0$. Any scattering involving $u(q)\ne 0$ will yield phonons in a harmonic system, and is accounted for by the higher order "moments" in (\ref{eq-Sqw1}).
As we will justify below, $S(q)$ may write as a sum of two terms:
\begin{eqnarray}\label{eq-Sb2} 
& S(q) &
=  \frac{1}{2 \pi \hbar  }  e^{-2W} \sum_{\l,\l'} 
e^{\imath \qb\cdot [\Rb_\l(t)-\Rb_{\l'}(0)]}  
 \int       e^{-\imath \w t}d t \lf[ f_0 e^{\imath 0   } + f_b e^{\imath 2 \pi  } \rt]   \qquad \quad({\rm a}) \qquad  
\nonumber \\
&& =  S_0^\el(q) +  S^\inel_b(q)     \qquad \qquad  \qquad  \qquad  \qquad  \qquad  \qquad\qquad  \qquad  \quad  ({\rm b})  
\end{eqnarray} 
Where $W$ is Debye-Waller factor;  $\Rb_l(t)$ is the equilibrium position of  nucleus $l$ at time $t$; the nucleus' displacement from $\Rb_l$ is $\ub_l$.
 The "partial" structure factors, $S_0^\el(q)$ and $S^\inel_b(q)$, describe an elastic and inelastic scattering, respectively, 
with the  corresponding scattered neutrons undergoing a phase change of zero and of $2\pi$.  
The subscripts $0$ and $b$ 
indicate  the momentum changes the incident neutron undergo are  $0$ and $q_b$. 
  $f_0$ and $f_b$ measure the fractions of the two kinds of scattering events, and satisfy  $f_0+f_b=1$. 
The first terms usually presents with harmonic solids, whilst the second term is specific with the superfluid scattering. 


\section{
Elastic scattering of neutrons about $q_b$}\label{Sec-DynStr-Elas}
\index{Structure factor, due to elastic scattering}
$S_0^\el(q)$ of (\ref{eq-Sb2}b) corresponds to the first summation term of (\ref{eq-Sb2}a):
\begin{eqnarray}\label{eq-Soel1}
S_0^\el(q) =  \frac{1}{2 \pi \hbar  }  e^{-2W} \sum_{\l,\l'} 
e^{\imath \qb\cdot [\Rb_\l(t) -\Rb_{\l'}(0)]}  
 f_0 \int       e^{-\imath \w t}d t  e^{\imath  0   } 
\end{eqnarray}
 (\ref{eq-Soel1}) describes an elastic  scattering of the neutron, i.e. the momentum of the neutron upon scattering may undergo a change in direction only but not in magnitude. 
The superfluid atoms are accordingly not produced with any displacement or energy change. 
The associated phase change on the time axis is thus zero, as is represented by the exponential term $e^{ \imath 0}$. 

As is contrasted to  a normal liquid which does not give truly elastic scattering except at $q=0$, the superfluid has the scheme for causing elastic scattering. This is because the energy levels of the superfluid are quantized, as can be satisfactorily accounted for by a SHM (simple harmonic motion)-RSB (relative to superfluid block) dynamics scheme\cite{Sup2}. A momentum transfer will thus not be accompanied with an energy transfer (the case of elastic scattering) unless it equals exactly the energy of a phonon.  
By contrast, in a normal liquid, the atoms can assume an energy over a continuous range at a given $q$, so a momentum transfer can always be accompanied by an energy transfer.  

For a disordered system, the superfluid here, the site summation
in (\ref{eq-Soel1}), and similarly in (\ref{eq-Sb3}) below, has the well-known result 
\begin{eqnarray} \label{eq-sum1} 
 \sum_{\l, \l'} e^{\imath \qb\cdot [\Rb_\l(t)-\Rb_{\l'}(0)]}  
= N [ 1+ \int g(R) e^{\imath \qb\cdot \Rb  } d R]
\end{eqnarray}
where $\Rb 
=\Rb_\l(t)-\Rb_{\l'}(0)$.
The time integration in (\ref{eq-Soel1}) yields 
\begin{eqnarray}\label{eq-Sqoelt}
\int   e^{-\imath (\w-0) t} \ d t =\delta(\w-0)     
\end{eqnarray}
Substituting (\ref{eq-sum1}) and (\ref{eq-Sqoelt}) into (\ref{eq-Soel1}), we have the explicit expression  
$$
S_0^\el(q)=   \frac{N}{2\pi \hbar}e^{-2W} \lf[ 1+ \int g(R) e^{\imath \qb\cdot \Rb  } d R\rt] f_0 \delta(\w-0)      
\eqno(\ref{eq-Soel1})'
$$
In the elastic scattering here, the neutrons do not directly probe the many-quantum-atom correlation in the superfluid. Instead the neutrons will see the instantaneous atomic configurations in the liquid,  dominated by two factors: 
(1) the  equilibrium positions of the superfluid are short-range disordered, and 
(2) at each moment in time the atoms are thermally irregularly  displaced to instantaneous positions from their equilibrium positions. The thermal displacement of the helium atoms is  particularly large owing to their small mass; this is reflected by a large exponent $W$ in the Debye-Waller factor in  (\ref{eq-Soel1})$'$. 
The particularly large atomic displacement in He II, which does not present in normal fluids and harmonic solids with larger atomic masses, determines a particularly broad peak in $S_0^{\el}(q)$, and subsequently in $S(q)$, of the superfluid. 
This provides an explanation why the structure factor of the superfluid, as revealed from scattering  measurements, is abnormally smeared (see e.g. \cite{Sup2}), 

 The atomic bonding energy of a solid is typically  in the range 1--10 eV, which is much greater than the thermal neutron energy $E_\neu$ (5 -- 100 meV). Thus the atomic bonds cannot be broken up (or excited) by the impingement of a thermal neutron.
This process is described by $S_b^{\inel}(q)$ to be discussed in Sec. \ref{Sec-DynStr-Ielas}.  
Hence, for a harmonic solid, $S_b^{\inel}(q)=0$;  so $S_0^\el(q)$ given in (\ref{eq-Sqw1}) is practically the only term in the structure factor of (\ref{eq-Sb2}), that is:
$$ S^{\rm solid}(q)= S_0^\el(q).
                                 \eqno(\ref{eq-Sb2}{\rm a})$$
If the solid is also crystalline, then (\ref{eq-Sb2}a) describes a Bragg scattering, i.e. $S^{\rm solid}(q) \propto \delta(q-q_b)$. 

\section{Inelastic scattering at $q_b$. The superfluid bond excitation} 
\label{Sec-DynStr-Ielas}
For the superfluid, the atomic bonding energy, being  $-7.2 $ K/atom ($ 0.62$ meV), is  $<<E_\neu$. 
The superfluid bond can therefore be easily broken up by the impingement of a thermal neutron; one thus expects inelastic neutron scattering to occur at $q_b$, corresponding to the excitation of the superfluid bond. We actually obtained this through the solution of equation of motion in  Ref. \cite{Sup2}. 
The theoretical anticipation  agrees with observation from the thermal neutron measurement,
namely that  at $q_b$ the neutron energy transfer is finite, and is $8.6$ K. 

Given that an excitation has occurred and the particle wave has evolved with time,   
the phase change on the time axis  will be non-zero, and this,  
as we will clarify below, is represented by the second partial structure factor in (\ref{eq-Sb2}). That is:
\begin{eqnarray} \label{eq-Sb3}
 S^\inel_b(q)   =  \frac{1}{2 \pi \hbar  }  e^{-2W_q} \sum_{\l,\l'} 
e^{\imath \qb\cdot [\Rb_\l(t)-\Rb_{\l'}(0)]}  
f_b \int     e^{\imath  2 \pi  }     e^{-\imath \w t}d t   
\end{eqnarray}
We below express the respective terms of (\ref{eq-Sb3}) explicitly for the present inelastic scattering at $q_b$, and prove that the phase factor $ e^{\imath  2 \pi  } $ results from the superfluid bond excitation.

Firstly, the Debye-Waller factor  at $q_b$ may be evaluated ordinarily 
to be: 
   \begin{eqnarray}\label{eq-Wq}
 W_q|_{q=q_b}=\frac{1}{2} q_b^2 <u>^2 =0.
   \end{eqnarray}

\index{Bragg scattering, inelastic}\index{Inelastic Bragg scattering}
We next carry out the site summation.
 As just recalled, the neutron scattering at $q_b$ involves an energy transfer, $\Ds$. In such a scattering process  a neutron directly communicates with---or probes---the many-quantum-atom correlation in the superfluid. The many-body nature  
leads to an excellent averaging effect to the fluctuation in cage size and atomic bonding strength in the superfluid over time and locations,  yielding an average fluid structure as  effectively seen by the scattered neutron. This hence structurally prepares for the neutron scattering intensity to satisfy at $q_b$  the Bragg condition: \index{Bragg scattering, inelastic}\index{Inelastic Bragg scattering} 
\begin{eqnarray} \label{eq-qR}
q_b\cdot [\Rb_{\l}(t)-\Rb_{\l'}(0)]=q_b\cdot \Rb = q_b n a = n 2\pi,  \qquad n=0, 1, \ldots   
\end{eqnarray}
where 
$a$ is the apparent interatomic spacing defined by $a=2\pi/q_b$.
It can be readily justified that \cite{Sup2}
the Bragg condition (\ref{eq-qR}) indeed holds to a high degree of approximation for the superfluid also according to the relation for neutron's wavevectors before and after scattering, which owing to the inelastic scattering suffers a change that is however negligibly small.  
With (\ref{eq-qR}), the site summation  
in (\ref{eq-Sb3})  writes:
\begin{eqnarray} \label{eq-RRp}
 \sum_{\l,\l'} 
e^{\imath \qb\cdot [\Rb_\l(t)-\Rb_{\l'}(0)]} = \delta(q-q_b). 
\end{eqnarray}
\index{Bragg scattering, inelastic}\index{Inelastic Bragg scattering}

We lastly derive the phase factor $ e^{\imath  2 \pi  } $ of (\ref{eq-Sb3}).  The $N$-particle system before and after a scattering event 
is in terms of its macroscopic property unchanged, despite the local perturbation taking place.  
The total wave functions of the $N$ identical bosonic particles 
(which solution is given in \cite{Sup2})
before and after the scattering must therefore be the same. For the $N$ boson particle system, this implies that the total phase change of the wave function  due to the superfluid bond excitation  must satisfy: 
\begin{eqnarray}\label{eq-tot-ph-ch} {\rm total \ phase \ change = n' 2\pi} \end{eqnarray}
$n'$ being integer.
The total phase change is a consequence of the evolution of the scattering system in both space and time.  
As shown by (\ref{eq-qR}), however, the phase change due to the evolution along the $X$ axis, i.e. $q_b a$, alone satisfies $2\pi$. Then, subtracting  (\ref{eq-qR}) from (\ref{eq-tot-ph-ch}) gives that the phase change associated with time evolution must alone also satisfy integer times $2\pi $, denoting $n'' 2 \pi$. $n''=1$ gives the smallest finite phase change and corresponds to one-superfluid bond excitation: \index{Superfluid bond, partial structure factor} \index{Structure factor, due to superfluid bond excitation}
\begin{eqnarray} \label{eq-Dbt}
&& 0  \quad  {_{\rm   
{superfluid \ bond \atop \ activation}
}\atop \longrightarrow  }\quad     \frac{\Ds}{\hbar} \tau_b=  2 \pi, 
\end{eqnarray} 
The corresponding contribution to the scattering function is then 
\begin{eqnarray}\label{eq-Dyna-phas-fac}
e^0 \quad  {_{\rm   
{superfluid \ bond \atop \ activation}
}\atop \longrightarrow  }\quad      e^{ \frac{\Ds}{\hbar} \tau_b}=    e^{\imath 2\pi}
\end{eqnarray}
The presence of a phase factor $e^{\imath 2\pi}$ in (\ref{eq-Sb3}), as finally derives from (\ref{eq-Dyna-phas-fac}), is thus proven. 

Substituting (\ref{eq-Wq}),  (\ref{eq-RRp}) and (\ref{eq-Dyna-phas-fac}) into (\ref{eq-Sb3}), 
we obtain the explicit expression for the partial structure factor of the superfluid due to the inelastic scattering of neutrons at $q_b$, upon the creations of one-superfluid bonds:
 $$\displaylines{
\quad\qquad \qquad \qquad \qquad \qquad S^\inel_b(q) 
=  \frac{1}{2 \pi \hbar N } \delta(q-q_b) 
f_b   \int^{\infty}_{-\infty}   e^{-\imath (\w-\frac{\Ds}{\hbar}  )t}\delta(\tau_b) d t  \hfill 
\cr
\qquad \qquad \qquad \qquad \qquad\qquad \qquad \ \ =   \frac{1}{2 \pi \hbar N }   \delta(q-q_b) f_b \delta(\w-\frac{\Ds}{\hbar})    \hfill (\ref{eq-Sb3})'
}$$
Since for the given $q=q_b$, $\w (q)= \Ds/\hbar$. Thus 
the function $S^\inel_b(q)$ at $q_b$ also represents the scattering function at a single point on the $\w$ axis. That is,  
\begin{eqnarray} \label{eq-SbDs}
S^\inel_b(q)|_{q=q_b}= S_b(q, \w)|_{\w=\frac{\Ds}{\hbar}}
\end{eqnarray}   
$S_b(q, \w)|_{\w=\frac{\Ds}{\hbar}}$ represents $S(q,\w)$ in the vicinity of $(q_b, \Ds)$; we shall not elaborate on $S(q,\w)$ in this article.  
 
The excitation energy $\w=\Ds$ at $q_b$ 
is basically well-defined, or single-valued  
 as a result that $\Db$ is single-valued 
due to  the many-quantum-atom correlation in the superfluid as discussed earlier. It follows that the peak in 
$S^\inel_b(q, \w)|_{\w=\frac{\Ds}{\hbar}}$ will be qualitatively sharp. This provides an explanation of the qualitatively sharp peak feature in  $S(q, \w)$ vs. $\w$ at $q=q_b$,  as observed in the inelastic neutron scattering experiments; an actually  finite broadening in the peak at $q_b$ can be argued  
attributable to the finite spreading in $\ds$.

\section{The total structure factor of the superfluid }
\index{Structure factor, total}

Substituting (\ref{eq-Soel1})$'$ 
and (\ref{eq-Sb3})$'$ into  (\ref{eq-Sb2}) 
we finally have the structure factor of the superfluid 
$$    S(q)=  f_0N \lf[ 1+ \int g(R) e^{\imath \qb\cdot \Rb  } d R\rt] \delta(\w-0)  + f_b \frac{1}{2 \pi \hbar N }  \delta(q-q_b)  \delta(\w-\frac{\Ds}{\hbar})  \eqno(\ref{eq-Sb2})'  $$
Since the two component functions of  (\ref{eq-Sb2}), $S_0^\el(q) $ and $ S^\inel_b(q)$ as explicated in (\ref{eq-Sb2})$'$,  represent two  qualitatively distinct scattering processes, their peak positions do not necessarily coincide. 
On the $q$ axis, as we surveyed in \cite{Sup2}
elastic neutron measurements directly show a peak in $S(q)$  positioned at $q_b'=2$ \AA$^{-1}$, corresponding to a reciprocal $a'=2$ \AA.
From the theoretical representation above  we see that this peak is predominately correlated to the elastic scattering process underlying  $S_0^\el(q) $, i.e. it corresponds to the peak in $S_0^\el(q) $. 
The maximum of $ S^\inel_b(q)$ is not directly revealed in the  measured $S(q)$ v.s $q$ curve. 
However, on the  $\w(q_b)$ axis  inelastic scattering experiments show a minimum of  
$S(q,\w)$   positioned  at $\w(q_b)$. 
If fixing $\w=\w(q_b)$ and plotting the function $S_b(q,\w(q_b))= S^\inel_b(q)$ on the $q$ axis, one would  expect a peak position at $q_b  =1.93$ \AA$^{-1}$, which does not coincide with $q_b'$. This peak at $q_b$ is presumably owing to its weak intensity not directly visible in the measured curve $S(q)$.

\paragraph*{Acknowledgements:} The core content of the new theory\cite{Sup2,Sup1} of superfluid helium  was developed by the author as a visiting scientist at the H.H. Wills Lab., Bristol Univ., 1998-1999,  work funded  by the Swedish Natural Science Research Council (NFR) and partly by the Wallenbergs Stiftelse (WS).  The author was indebted to the encouragement and stimulating discussions from a number of colleagues and intentional specialists in the related fields.

\end{document}